\documentclass[preprint2,numberedappendix]{emulateapj-rtx4}
\usepackage{natbib}
\usepackage{graphicx,times,bm,url}
\usepackage{afterpage,lscape}
\graphicspath{{./fig/}{./png/}}

\newcommand{\EQ}{\begin{equation}}
\newcommand{\EN}{\end{equation}}
\newcommand{\EQA}{\begin{eqnarray}}
\newcommand{\ENA}{\end{eqnarray}}

\newcommand{\Fig}[1]{Figure~\ref{#1}}

\newcommand{\Figs}[2]{Figures~\ref{#1} and \ref{#2}}

{}
{}
{}

{}
{}
{}
{}
{}
{}
{}
{}
{}
{}
{}
{}
{}
{}
{}
{}
{}
{}
{}
{}
\newcommand{\meanUU}{\overline{\mbox{\boldmath $U$}}{}}{}

{}
{}
{}

{}

{}
{}

\newcommand{\BB}{\bm{B}}

\newcommand{\uu}{\mbox{\boldmath $u$} {}}

\newcommand{\JJ}{\mbox{\boldmath $J$} {}}

\newcommand{\nab}{\mbox{\boldmath $\nabla$} {}}

\newcommand{\oo}{\mbox{\boldmath $\omega$} {}}

\def\Ro{\mbox{\rm Ro}}

\def\Sh{\mbox{\rm Sh}}

\def\Teff{T_{\rm eff}}

\def\kf{k_{\rm f}}

\def\urms{u_{\rm rms}}

\def\etat{\eta_{\it t}}

\def\Beq{B_{\rm eq}}

\newcommand{\K}{\,{\rm K}}

\newcommand{\erg}{\,{\rm erg}}

\usepackage{amsmath}
\begin{document}

\title{Superflare occurrence and energies on G, K and M type dwarfs}

\author{S. Candelaresi$^{1,2,3}$}
\author{A. Hillier$^{4}$}
\author{H. Maehara$^{4,5}$}
\author{A. Brandenburg$^{2,3}$}
\author{K. Shibata$^{4}$}
\affil{
$^1$Division of Mathematics, University of Dundee, Dundee, DD1 4HN, UK \\
$^2$Nordita, KTH Royal Institute of Technology and Stockholm
University, Roslagstullsbacken 23,
SE-10691 Stockholm, Sweden \\
$^3$Department of Astronomy, AlbaNova University Center,
Stockholm University, SE-10691 Stockholm, Sweden \\
$^4$Kwasan and Hida Observatory, Kyoto University, Yamashina, Kyoto 607-8471, Japan \\
$^5$Kiso Observatory, Institute of Astronomy, School of Science,
The University of Tokyo 10762-30, Mitake, Kiso-machi, Kiso-gun, Nagano 397-0101, Japan
}


\received{2014 May 6}
\accepted{2014 July 3}
\published{2014 August 18}
\journalinfo{The Astrophysical Journal, 792:67 (9pp), 2014 September 1
\hfill doi:10.1088/0004-637X/792/1/67}

\begin{abstract}
\textit{Kepler} data from G-, K-, and M-type stars are used to study conditions that lead
to superflares with energies above $10^{34} {\rm erg}$.
From the 117,661 stars included, 380 show superflares with a total of
1690 such events.
We study whether parameters, like effective temperature or rotation rate,
have any effect on the superflare occurrence rate or energy.
With increasing effective temperature we observe a decrease
in the superflare rate, which is analogous to the previous findings of a
decrease in dynamo activity with increasing effective temperature.
For slowly rotating stars, we find a quadratic increase of the mean occurrence rate
with the rotation rate
up to a critical point, after which the rate decreases linearly.
Motivated by standard dynamo theory, we study the behavior of the relative
starspot coverage, approximated as the relative brightness variation.
For faster rotating stars, an increased fraction of stars shows
higher spot coverage, which leads to higher superflare rates.
A turbulent dynamo is used to study the dependence of the Ohmic dissipation
as a proxy of the flare energy on the differential rotation or shear rate.
The resulting statistics of the dissipation energy as a function of dynamo
number is similar to the observed flare statistics as a function of the
inverse Rossby number and shows similarly strong fluctuations.
This supports the idea that superflares might well be possible for
solar-type G stars.
\end{abstract}

\section{Introduction}

Research in solar and stellar variability has often focused on
grand minima, but seldom on grand maxima.
One of the characteristics of a grand maximum may be an enhanced frequency
of superflares.
Superflares release energies of $10^{34}\erg$ or more.
Such flares are not generally expected to occur in the Sun, where the
strongest flares have only about $10^{32}\erg$; an example of this is
the Carrington flare of 1859 \citep{Carrington1859MNRAS,Hodgson1859MNRAS}.

Exhaustive statistics of superflares in other solar-like stars with
solar rotation rates
\citep{Maehara2012Natur,Shibayama-Maehara-2013-209-1-ApJS,Nogami-Notsu-2014-1402-arxiv}
have reinvigorated the discussion
of whether or not such events could in principle also occur in the Sun.
Of particular importance is the realization that ``hot Jupiters'' are {\em not}
required \citep{Shibata2012,Shibayama-Maehara-2013-209-1-ApJS},
contrary to what was previously believed
\citep{Schaefer2000ApJ}.
The recent work of \cite{Shibata2012} explored the possible connection between
flare intensity and sunspot area.
They argue that flux transport dynamos
\citep{Choudhuri1995AA,Dikpati1999ApJ,Nandy2002Science} might be
capable of generating enough magnetic flux, storing
it beneath the convection zone for some time, and then releasing it in a
violent eruption.
One of the aims of the present paper is to
find favorable conditions under which superflares can occur.
We will
discuss an alternative scenario for the origin of
superflares within the framework of turbulent dynamo theory.

Dynamo activity is connected to stellar rotation through the dependence
of the $\alpha$ effect and shear on the rotation rate.
Both quantities are important ingredients in large-scale dynamos,
where the $\alpha$ effect is necessary in most dynamo models
\citep{steenbecketal66,PouquetFrischLeorat1976JFM,BrandenbSubramanianReview2005}.
A positive correlation between the star's rotation rate and the occurrence
of superflares is therefore expected.

It is generally believed that
as the star's magnetic energy increases, more and larger starspots can occur.
We expect that this excess magnetic energy is stored until
the spot dissolves or magnetic reconnection initiates a flare
\citep{Su-Veronig-2013-9-489-NatPh,Malanushenko-Schrijver-2014-783-2-ApJ}.
Individual flare events can only be resolved for the Sun.
For distant stars, they need to be inferred by characteristic brightness
variations.
\cite{Maehara2012Natur} found 148 stars with 365 such events by searching
for peaks in the light curves from data compiled by the \textit{Kepler} mission
\citep{KochKeplerMission2010}.
Subsequent work by the same group \citep{Shibayama-Maehara-2013-209-1-ApJS}
extended the number of stars to 279 and flare events to 1547.

Starspots are manifested through periodic variations of a star's
luminosity.
From past work we know that inferring spot coverage or spot size from light
curves is accompanied by large uncertainties
\citep{Kovari-Bartus-1997-323-801-AA}.
Therefore, we perform calculations for the brightness variations of model stars
for which we know the spot distribution and measure the statistical spread.

Dynamo theory describes the conversion of kinetic energy into magnetic
energy.
In late-type stars such as the Sun, the kinetic energy comes from
convection in the outer layers.
Since flares are associated with magnetic fields, it seems clear that
the cause of superflares should be explicable in terms of dynamo theory.
Dynamo theory is a broad subject encompassing both small-scale and
large-scale dynamos.
Usually, only the large-scale dynamo is associated with the solar cycle,
but small-scale dynamo action might well occur at the same time.
In fact, at small scales, the two cannot even be distinguished,
because their magnetic and kinetic power spectra are virtually the same
\citep{Brandenburg2012SSR}.

It is conceivable that flare activity is more directly related to
the small-scale part of turbulence.
Indeed, only at small length scales does hydrodynamic and hydromagnetic
turbulence display the characteristics of strong intermittency with
bursts and long waiting times \citep{Veltri2005NPGeo}, required to explain
a broad range of different flares, including superflares.
On the other hand, both flares and coronal mass ejections may also be associated
with magnetic helicity \citep{Schrijver2009AdSpR43}, whose long-term variability
is certainly a feature of large-scale dynamos.
Superflares might therefore be the result of the simultaneous occurrence
of two or more time-dependent stochastic events.

\section{Flare Activity, Rotation, and Temperature}

EUV images of the Sun have long revealed the presence of magnetically
confined hot plasma.
Thermal X-ray emission from such hot plasma therefore provides proxies of magnetic activity
\citep{Pallavicini1981,Walter1981,Vilhu1984AA}
which are all correlated with the Coriolis or inverse Rossby number,
$\Ro^{-1}=\tau/P_{\rm rot}$, where $P_{\rm rot}$ is the rotation period and
$\tau$ is the convective turnover time.
To check whether superflare activity also correlates with $\Ro^{-1}$,
we consider the set of superflare stars identified by \cite{Maehara2012Natur},
which was subsequently extended to quarters 0 to 6 of the \textit{Kepler} survey
\citep{KochKeplerMission2010} with
380 superflare stars, 373 of which
have a well-determined rotation period.

We consider the superflare frequency $\nu$, which
is the number of superflares per unit time.
Here, a superflare is defined as an event that releases
a total energy of $5\times10^{34}\erg$ or more within a few hours.
Of the updated sample of \cite{Maehara2012Natur}, there are 129 G-type stars
with effective temperatures in the range $5200\K\leq\Teff<6000\K$,
227 K-type stars with $3700\K\leq\Teff<5200\K$, and
17 M-type stars with $2400\K\leq\Teff<3700\K$.
To determine their Rossby numbers, we use the empirically determined
turnover times $\tau$ of \cite{Noyes1984ApJ},
who found a relation between $\tau$ and the $B-V$ color.

\begin{figure}[t!]\begin{center}
\includegraphics[width=\columnwidth]{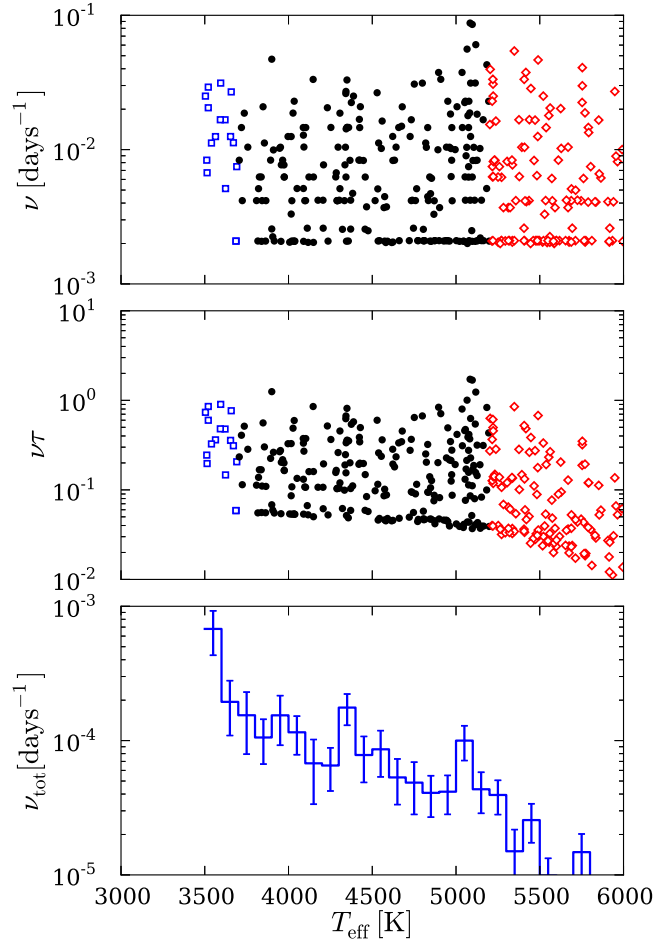}
\end{center}\caption[]{
Unnormalized flare frequency $\nu$ vs.\ $\Teff$ (upper panel),
normalized $\nu\tau$ (central panel) and binned averages for the 
flare frequency $\nu_{\rm tot}$ including non-superflaring stars
(lower panel).
Blue squares refer to M-type stars (effective temperatures below $T_1=3700\K$)
and red diamonds to G-type stars (temperatures above $T_2=5200\K$).
Black dots refer to K-type stars (temperatures between $T_1$ and $T_2$). 
The lower cutoff in $\nu$ is $\nu_{\min}=1/t_{\max}$ and
is an artifact of the finite length of the observed time series, $t_{\max}$.
}\label{fig: nuTau_vs_Teff}\end{figure}

\begin{figure}[t!]\begin{center}
\includegraphics[width=\columnwidth]{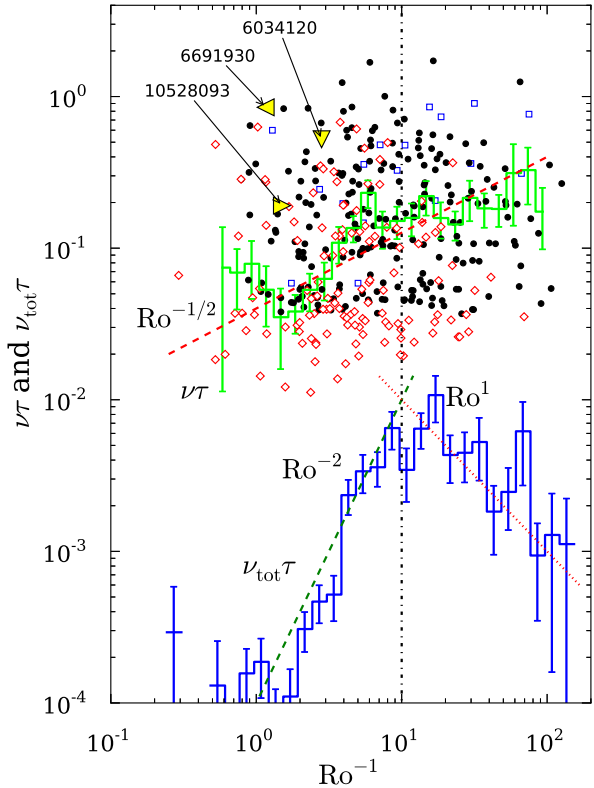}
\end{center}\caption[]{
Non-dimensional flare frequency $\nu\tau$ vs.\ $\Ro^{-1}=\tau/P_{\rm rot}$.
Color and shape coding of the markers are the same as in
\Fig{fig: nuTau_vs_Teff}.
The stars considered in \cite{Notsu2013ApJ} are highlighted as yellow triangles,
together with their \textit{Kepler} ID.
The upper (green) curve shows the averaged superflare frequencies for intervals
in $\Ro^{-1}$ including only superflaring stars.
From the seemingly random scatter plot and the flat profile for the averaged
$\nu\tau$, it becomes clear that there is no correlation between $\Ro^{-1}$
and $\nu\tau$.
In contrast, the lower (blue) plot shows the average value of $\nu_{\rm tot}\tau$
(which includes non-superflaring stars) as a function of the inverse Rossby number.
There, two regimes can be distinguished.
For $\Ro^{-1} \lesssim 10$, we find a power law of $-2$, while for higher values we find
$1$ for the power.
}\label{fig: nuTau_vs_Ro-1_mixed}\end{figure}

In \Fig{fig: nuTau_vs_Teff}, we plot the dimensional and non-dimensional
superflare frequencies, $\nu$ and $\nu\tau$ versus $\Teff$.
We see that the superflare rate is uniformly distributed
and nearly independent of the star's effective temperature.
We further plot $\nu\tau$ versus $\Ro^{-1}$
(\Fig{fig: nuTau_vs_Ro-1_mixed}).
It turns out that there is no clear correlation between
$\nu\tau$ and $\Ro^{-1}$, as can also be seen from the nearly flat profile
of the green line in \Fig{fig: nuTau_vs_Ro-1_mixed}, which shows the
average taken over all superflaring stars within the shown interval
in ${\rm Ro}^{-1}$.

Note that the analysis above only includes stars which showed superflares
during the observation period.
Such an analysis does not give any hint as to whether or not all stars with
specific parameters, like the Rossby number, are more likely to produce superflares.
By normalizing with respect to all observed \textit{Kepler} stars from quarters 0 to 6,
we compute the average superflare occurrence rate, $\nu_{\rm tot}$, over a set of
binning intervals where we include both superflaring and non-superflaring stars
(the total number of those stars is 117,661, only 115,984 of which with a well defined
rotation period were included).
We find a clear decrease of the superflare frequency $\nu_{\rm tot}$ with
increasing effective temperature (\Fig{fig: nuTau_vs_Teff}, lower panel).
This might seem to be in contradiction with observational results
showing a clear positive power-law dependence of the surface shear on the
effective temperature \citep{Barnes-CollierCameron-2005-357-L1-MNRAS}.
Subsequent model calculations, however, showed that for hotter stars, eddy
diffusion increases such that it overcompensates for the increase in shear
which leads to a reduced dynamo number
\citep{Kitchatinov-Olemskoy-2011-411-2-MNRAS}, thus explaining our results.

For the Rossby number dependence, we find two regimes separated by a boundary at
$\Ro^{-1} \approx 10$.
For lower values, the superflare occurrence rate shows a power law proportional
to $\Ro^{-2}$, while for higher $\Ro^{-1}$ we find a $\Ro^{1}$ behavior
(\Fig{fig: nuTau_vs_Ro-1_mixed}, lower curve).
With a total of 373 superflaring stars, we have confidence in the
validity of our statistical analysis.
Furthermore, the determined power laws fit the computed averages remarkably well.

The position of the break at $\Ro^{-1}\approx10$ agrees with the
well-known point of saturation of chromospheric and X-ray activity
for large values of $\Ro^{-1}$ \citep[see, e.g.][]{Pizzolato2003,Wright2011ApJ}.
For smaller values of $\Ro^{-1}$, they find an approximately
quadratic increase of X-ray luminosity, which is similar to our
quadratic increase of flare frequency.
At larger values of $\Ro^{-1}$, stellar activity is saturated
and thus not compatible with the fall-off seen in the lower curve of
\Fig{fig: nuTau_vs_Ro-1_mixed}.

\section{Relation to Starspots}

\subsection{Relative Star Population}

Enhanced magnetic activity is manifested through the increased occurrence of
starspots.
Starspots can be inferred through cyclic variations in the light curves
with frequencies equal to the rotation frequency.
For every star, we know its relative flux variation $\Delta F/F_{\rm av}$
\citep{Maehara2012Natur},
where $\Delta F$ is the range over which the flux varies
and $F_{\rm av}$ is the averaged flux.
According to \cite{KochKeplerMission2010}, the shot noise for stars measured
over 6.5 hr is 14 ppm, which will be our detection limit for
$\Delta F/F_{\rm av}$.
We use this as proxy
for the fraction of the stellar surface covered by spots.
In principle, there can be other causes for the observed flux variation,
e.g., differential rotation \citep{Reinhold-Reiners-2013-560-A4-AA} and
exoplanets, but the latter were excluded by \cite{Maehara2012Natur}.
Furthermore, starspots which are visible during a whole revolution,
like those spots extending over the poles, would weaken the
applicability of the flux variation as a proxy for starspot coverage,
since they would constantly reduce the flux.
A thorough study of the observational bias is presented in Section
\ref{sec: spot modeling}.

\begin{figure}[t!]\begin{center}
\includegraphics[width=\columnwidth]{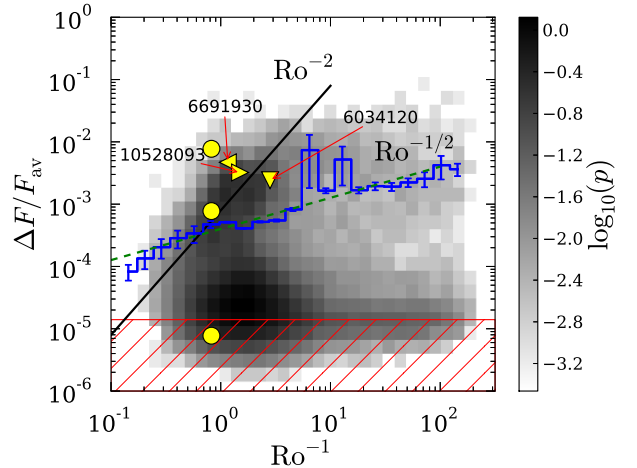}
\end{center}\caption[]{
Average value of $\Delta F/F_{\rm av}$ dependent on the Rossby number including
all observed 115,984 stars (1D histogram).
Probability distribution function for the same set of stars dependent
on $\Delta F/F_{\rm av}$ and $\Ro^{-1}$ (color mapping).
The hatched area indicates the detection limit.
Standard dynamo predictions for which magnetic activity is expected to
increase with rotation rate \citep{Karak2014ApJ}
are confirmed for this set of stars,
provided the relative flux variation is a good proxy for the total
magnetic flux.
Apart from a general increase of the relative brightness variation, we
observe two regimes (darker shades).
The upper regime can be identified as the active regime where the stars were
magnetically
active during the observation.
Here, one observes an increase of the flux variation as the rotation rate increases.
For the stars in their magnetic minima, rotation does not play any
role.
Three particular superflaring stars investigated in \cite{Notsu2013ApJ} are highlighted here as yellow
triangles together with their \textit{Kepler} ID.
The strong brightness variation of those three stars indicates large spots which
can trigger superflares, as seen in \Fig{fig: nuTau_vs_Ro-1_mixed}.
For comparison, we also indicate the Sun's position in this diagram if it was
covered with sunspots for 1\%, 0.1\%, and 0.001\% (upper, middle, and lower yellow
circle, respectively).
}\label{fig: Ro-1_df_hist_mixed}\end{figure}

We expect enhanced dynamo activity and larger and more starspots
for rapidly rotating stars, i.e., for large values of $\Ro^{-1}$.
This is reflected in the dependence of the relative flux variation for the
115,984 stars in this catalog (\Fig{fig: Ro-1_df_hist_mixed}, one-dimensional (1D) histogram),
where we plot the average of
$\Delta F/F_{\rm av}$ for certain intervals of $\Ro^{-1}$.
Apart from two outliers, we obtain
good agreement with a power-law dependence of
$\Delta F/F_{\rm av} \propto \Ro^{-1/2}$.
Analytical mean-field dynamo calculations and simulations for rotating shearing
dynamo models \citep{Karak2014ApJ} have shown that the
saturation magnetic field strength increases with the rotation frequency
$\Omega$ with a power law of $\Ro^{-1/2}$, which
is in agreement with the observations.

\textit{Kepler} observations cover stars at a random phase of their magnetic
activity cycle.
During low magnetic activity, very few starspots are expected, independent
of the Rossby number.
During high magnetic activity, strong flux variations are expected.
Those variations can depend well on the rotation rate, as dynamo activity
is expected to increase with $\Ro^{-1}$.
Those two branches of low and high magnetic activity
can be seen in \Fig{fig: Ro-1_df_hist_mixed} (color mapping), where we plot a
two-dimensional probability distribution for stars showing
a brightness variation and Rossby number in a certain interval.
The two regimes are clearly visible, where for one there is no
$\Ro$ dependence, while for the other we observe
$\Delta F/F_{\rm av} \propto \Ro^{-2}$ for
$\Ro^{-1} \lesssim 3.2$
and a constant for $\Ro^{-1} \gtrsim 3.2$.

\subsection{Sunspot Coverage Since 1874}

The Sun's cyclic variations and historic grand minima show that magnetic
activity can exhibit significant long-term variability.
There is no reason to attribute these characteristics exclusively to our
nearest star.
Other stars may have been in phases of high and low magnetic activity during
\textit{Kepler}'s observations.
The \textit{Kepler} mission, with its 90-500 days observing time, will have observed
stellar brightness at random phases, and hence starspot coverages.
This can explain the two regimes in \Fig{fig: Ro-1_df_hist_mixed}.
As the stars included here are all solar like, we should expect to
observe the Sun in either regime for a long enough observation interval.
Judging from \Fig{fig: Ro-1_df_hist_mixed}, there is a significant
chance that the Sun will be in the upper
or lower regimes, while the area in between is unlikely to be observable.

Using the data of
Hathaway\footnote{\url{http://solarscience.msfc.nasa.gov/greenwch.shtml}},
we know the sunspot coverage of the visible hemisphere
from 1874 May until 2013 July, with a time
resolution of one day and only a few short intervals with no observations.
From that data set, we compute the brightness variation.
The sunspot temperature is taken to be $T_{\rm s} = 4000\,{\rm K}$, while
$T_{\rm p} = 5800 {\rm K}$ is assumed for the photospheric temperature.
According to \cite{Notsu2013ApJ}, the brightness variation is calculated as
\EQ
\Delta F/F_{\rm av} = \left(1-(T_{\rm s}/T_{\rm p})^4\right) \times q,
\label{eq: dF_F} \EN
where $q$ is the relative sunspot coverage.
Any bright magnetic structures, which tend to exist on small scales,
would contribute to the
total magnetic energy while decreasing $\Delta F/F_{\rm av}$.
Since those are not included in the catalog for the Sun, we disregard
such structures in our calculations.
It has been found, however, that the Sun increases in irradiance by
approximately $1\%$ during high magnetic activity and high sunspot number
\citep{Frohlich-Lean-1998-25-23-GeoRL}.
This effect was attributed to facular brightening which overcompensates
the darkening effect from the sunspots.
As as faculae are much more
homogeneously distributed over the Sun, they do not lead to significant
brightness variations on timescales of the order of one rotation period.
As \textit{Kepler} observation times are much shorter than the magnetic cycle period,
the enhanced brightness appears as increased background radiation and does
not affect Equation \eqref{eq: dF_F}.
Variations in irradiance on timescales of days
\citep{Willson-Hudson-1981-244-L185-ApJL}, on the other hand,
were attributed to the appearance of
sunspot groups \citep{Willson-Gulkis-1981-211-700-Sci}, hence justifying our model.

\begin{figure}[t!]\begin{center}
\includegraphics[width=\columnwidth]{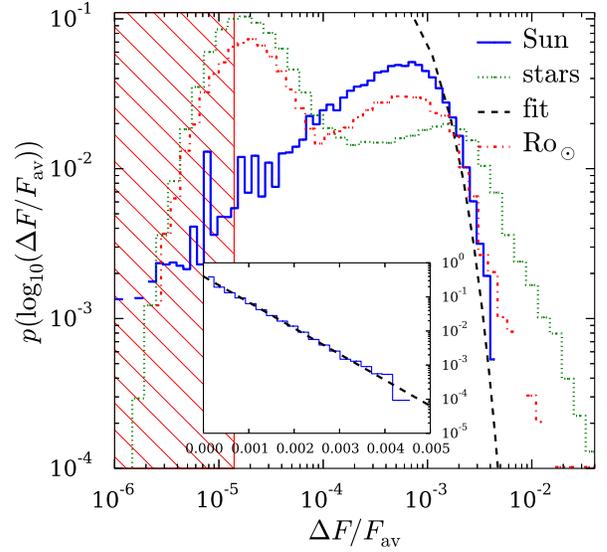}
\end{center}\caption[]{
Probability distribution function for observing a brightness variation
$\Delta F/F_{\rm av}$ at a random time between 1874 May and 2013 July
on the Sun (solid, blue line) for the 115,989 stars used in
\Fig{fig: Ro-1_df_hist_mixed} (dotted, green line) and stars with Rossby numbers
within $10\%$ of the solar value (dash-dotted, red line).
The inset shows the probability distribution $p(\Delta F/F_{\rm av})$
together with the fit.
We fit the Sun's data with the function $0.4\times \exp(-1750\Delta F/F_{\rm av})$
(dashed, black line).
The hatched area indicates the detection limit.
}\label{fig: dF_F_hist_sun_stars_daily}\end{figure}

The probability distribution function for finding the brightness variation
$\Delta F/F_{\rm av}$ at any time between 1874 May and 2013 July shows
an exponential shape (\Fig{fig: dF_F_hist_sun_stars_daily}), unlike the expected
two regimes with high probability for high and low brightness variation.
As the spot coverage of the superflaring stars is higher than
has ever been observed for the Sun, it is possible that,
given a long enough observing time of 5000 yr \citep{Shibata2012},
those two regimes might still appear.

\subsection{Superflares Related to Starspot Coverage}

\begin{figure}[t!]\begin{center}
\includegraphics[width=\columnwidth]{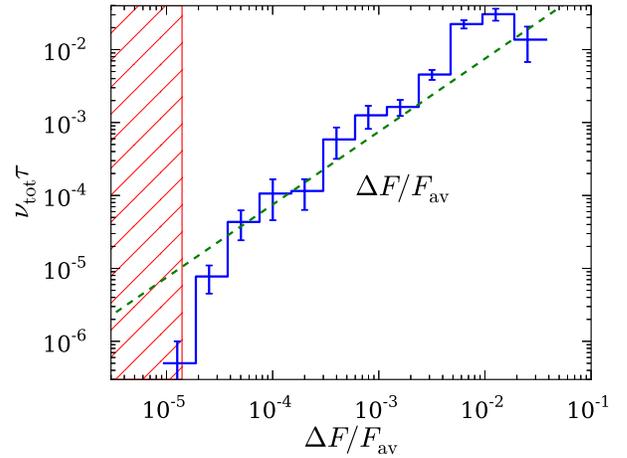}
\end{center}\caption[]{
Average superflare occurrence rate dependent on the relative brightness
variation including all the 115,984 stars observed.
The hatched area indicates the detection limit.
}\label{fig: nuTau vs dF}\end{figure}

Flares originate at the areas above starspots \citep{Sammis2000}
where magnetic field lines reconnect and give rise to particle acceleration.
Since we use the relative flux variation as a proxy for the area covered by
starspots, we expect more frequent superflares as $\Delta F/F_{\rm av}$
increases.
For the average superflare occurrence frequency, we determine an
approximate power law of $\nu_{\rm tot}\tau \propto \Delta F/F_{\rm av}$
(\Fig{fig: nuTau vs dF}).
Since we expect the flare area to increase with increasing $\Delta F/F_{\rm av}$,
we also expect the frequency of flares to increase with
the same power.

\begin{figure}[t!]\begin{center}
\includegraphics[width=\columnwidth]{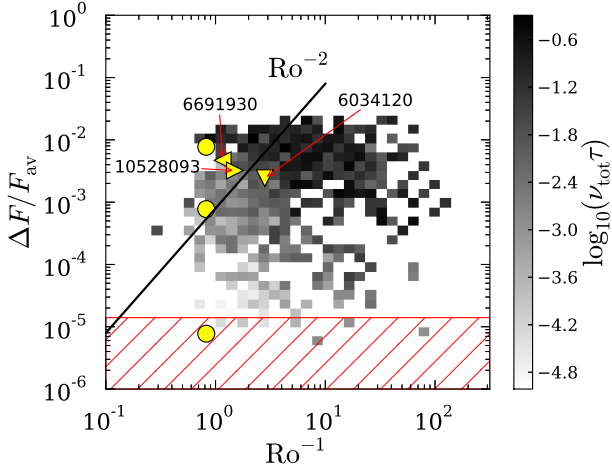}
\end{center}\caption[]{
Binned average for the superflare occurrence rate $\nu_{\rm tot}\tau$
dependent on the inverse Rossby number $\Ro^{-1}$ and the brightness
variation $\Delta F/F_{\rm av}$.
The yellow triangles and circles indicate the same stars, respectively, and our
Sun, as described in \Fig{fig: Ro-1_df_hist_mixed}.
The hatched area indicates the detection limit.
As there is only a weak dependence of the superflare rate on the inverse
Rossby number for constant $\Delta F/F_{\rm av}$, we can safely conclude
that we can observe superflares due to the appearance of starspots.
Those, in turn, are a product of fast rotation.
}\label{fig: nuTau_vs_Ro-1_df_all_binned}\end{figure}

From \Figs{fig: nuTau_vs_Ro-1_mixed}{fig: nuTau vs dF}, we can
conclude that an increase in the inverse Rossby number and/or starspot coverage
will lead to an increase in the superflare occurrence rate.
Increased brightness variation, however, is a consequence of
increased rotation rate, as shown in \Fig{fig: Ro-1_df_hist_mixed}.
The question to answer now is whether or not a star with strong brightness variation
will show high superflare rates regardless of the Rossby number.
To clarify this, we plot the binned average of the superflare occurrence rate
$\nu_{\rm tot}\tau$ for all the 115,984 stars as a function of both
$\Ro^{-1}$ and $\Delta F/F_{\rm av}$ (\Fig{fig: nuTau_vs_Ro-1_df_all_binned}).
This clearly shows a dependence of $\nu_{\rm tot}\tau$ on
$\Delta F/F_{\rm av}$, while the dependence on $\Ro^{-1}$ is
comparatively weak for fixed $\Delta F/F_{\rm av}$.
As an example, consider the horizontal ridge in
\Fig{fig: nuTau_vs_Ro-1_df_all_binned} through
$\Delta F/F_{\rm av}=10^{-2}$, along which the value of
$\nu_{\rm tot}\tau$ is nearly the same,
regardless of the value of $\Ro^{-1}$.
Hence, fast rotation leads to high starspot coverage, which increases the
chance for superflare eruptions.

\section{Starspot Modeling and Brightness Variation}\label{sec: spot modeling}

As alluded to earlier, we need to clarify to what extent
the brightness variation $\Delta F/F_{\rm av}$ can be used as proxy for
starspot coverage.
Since we cannot resolve the stars, we perform model calculations for the
brightness variation of stars covered with spots.
For each realization of the spot distribution, the flux variation during
one rotation is determined as a function of the star's inclination angle.
From that data, we determine the spread of the starspot coverage for given
ranges of $\Delta F/F_{\rm av}$.

A similar analysis was performed by \cite{Kovari-Bartus-1997-323-801-AA},
who reconstructed light curves from a synthetic measurement of an observed star
with 10 spots by using model stars with 2 spots.
To find the parameters of the two-spot model, they applied a minimization
technique which showed large uncertainties and a strong dependence on the
initial guess.
Since those fitting curves could approximate the synthetic light curve
very well, it could be concluded that there are strong ambiguities
in the reconstruction of such light curves.
Using a two-spot model, \cite{Notsu2013ApJ} reconstructed
observed light curves for stars at given inclinations.
On this basis, they concluded
that the brightness variation approximates the spot coverage well.
They did not, however, determine the ambiguity which arises from applying
different inclination angles and more than two spots.
In our starspot model, the aim is to make as few assumptions as possible and
to determine quantitatively the uncertainties.

\subsection{Model Stars}

We create a starspot map by randomly choosing the spot center on the
stellar surface in azimuth $\phi$ and latitude $\theta$, with the limits
$0 \le \phi < 2\pi$ and $-\pi/2 < \theta < \pi/2$.
For $\phi$ and $\theta$, random and uniformly distributed within their limits,
a higher density of spots is obtained near the poles.
This is automatically mitigated by choosing the spot size such that
$A_{\rm spot} \propto \cos\theta$.
Hence, the probability density function of spot coverage is a constant in
$\phi$ and $\theta$.
We know that for the Sun they appear more frequently close to the equator
at about $\pm30^{\circ}$ latitude.
We do not assume any such preferential distribution for our stars.
Furthermore, in our model stars, spots may extend over the equator
and the poles, thus allowing for a very general spot distribution.

\begin{figure}[t!]\begin{center}
\includegraphics[width=\columnwidth]{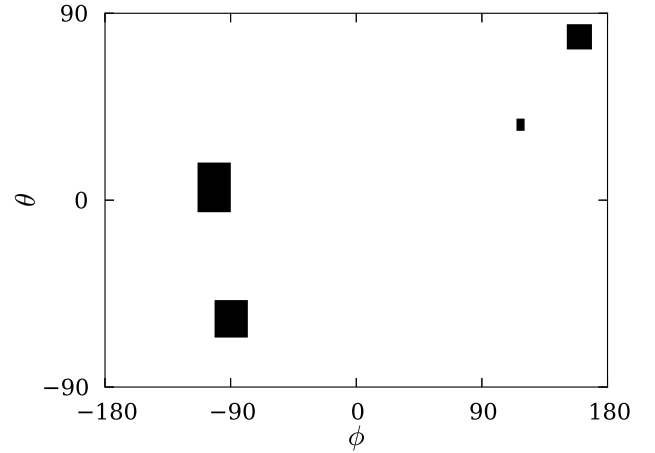}
\end{center}\caption[]{
Starspot map example used in the statistical analysis (equirectangular
projection).
The dark areas are those covered by spots and the white areas are
free of spots.
}\label{fig: spot_map_models}\end{figure}

For simplicity, the spots are taken to be segments of linear size $s$
extending from $\phi - s/2$
to $\phi + s/2$ and $\theta - s/2$ to $\theta + s/2$,
where $s$ is a random value between $0$ and $\pi/5$, thus covering
up to $3\%$ of the star's surface.
With a random number of spots between $0$ and $10$,
we can cover up to approximately a third of the surface.
Of course, starspots are not square-like,
but for our statistical analysis this is a good enough approximation.
An example spot coverage is plotted in \Fig{fig: spot_map_models}
where the dark areas indicate spots and the white areas are free of spots.

\subsection{Synthetic Observations}

From our starspot maps, such as \Fig{fig: spot_map_models}, we can extract
measurements of the observed brightness as a function of the inclination
angle $i$ of the observer to the equatorial plane and the facing
meridian $\phi$.
The latter is sampled through to simulate the star's rotation, which then gives rise
to a flux curve from which we extract $F_{\rm av}$ and $\Delta F$.
The total flux of the observed disk depends on the disk coverage with
spots and is simply the integral of the flux over the disk.
For the spot temperature, we use a solar value of
$T_{\rm s} = 4000\,{\rm K}$ and a photospheric temperature of
$T_{\rm p} = 5800\,{\rm K}$.
The emitted flux is proportional to $T^{4}$.
Since no information is available on the radial density and temperature
distribution of the stars, we neglect limb darkening effects.
Furthermore, we do not consider small-scale bright regions such
as plages.

\begin{figure}[t!]\begin{center}
\includegraphics[width=\columnwidth]{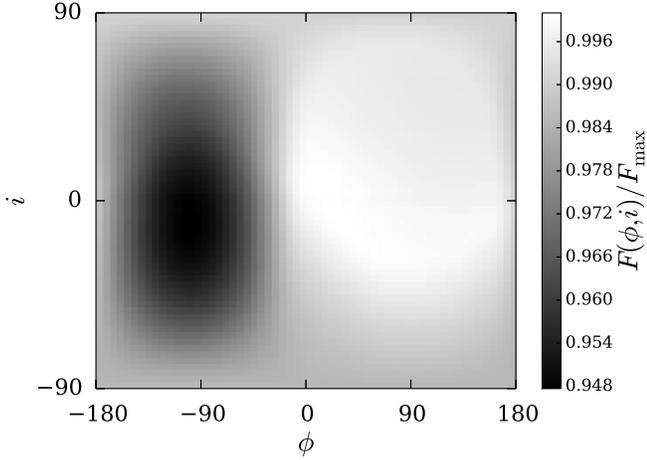}
\end{center}\caption[]{
Observed brightness of a star with spot distribution shown in
\Fig{fig: spot_map_models} dependent on the facing meridian
$\phi$ and inclination angle $i$ (equirectangular
projection).
}\label{fig: F_map_models}\end{figure}

We plot the observed flux dependent on $\phi$ and $i$ and obtain
the brightness map in \Fig{fig: F_map_models} for the previous example,
where $F(\phi,i)$ is the observed flux at a given inclination angle and
facing meridian, i.e., the rotation phase.
Comparing with \Fig{fig: spot_map_models}, we can readily confirm the validity of
the synthetic observations.

\subsection{Statistical Spread}

In order to obtain good statistics, we create 3771 realizations of starspot
distributions, each of which leads to a different starspot coverage
$A_{\rm spot}$.
Should the observed brightness variation be weakly dependent on the
particular realization or the star's inclination, we may conclude
that $\Delta F/F_{\rm av}$ is a good proxy for starspot coverage.

\begin{figure}[t!]\begin{center}
\includegraphics[width=\columnwidth]{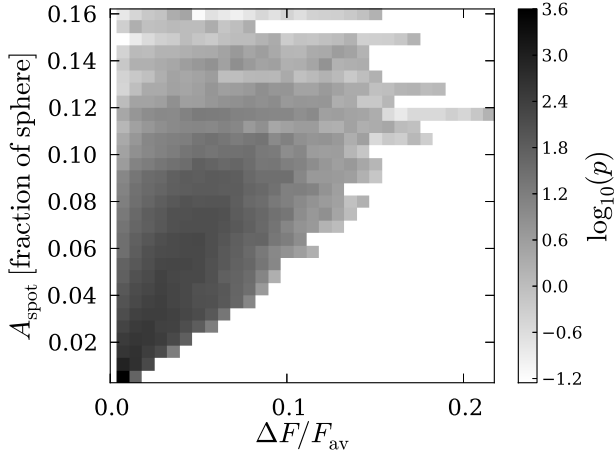}
\end{center}\caption[]{
Starspot coverage map from synthetic measurements for 3771 different
realizations of starspot coverage for various inclination angles
dependent on the brightness variation.
}\label{fig: A_dF_models_map}\end{figure}

In \Fig{fig: A_dF_models_map} we plot a map of the fractional spot coverage
for all inclinations and
realizations as a function of the brightness variation.
A general trend can readily be seen.
As the brightness variation increases,
the expected spot coverage also increases.
To determine the statistical significance, we bin the data for various
intervals of $\Delta F/F_{\rm av}$ and determine the mean for $A_{\rm spot}$.
Since high inclination angles are statistically less likely than face-on
observations, we need to weigh the significance of the data according to
the weight $\omega = \cos i$.

\begin{figure}[t!]\begin{center}
\includegraphics[width=\columnwidth]{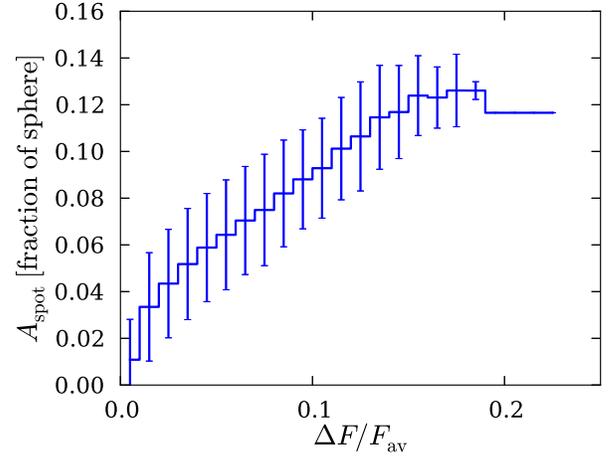}
\end{center}\caption[]{
Binned averages for the starspot coverage for intervals
of the observed brightness variation, together with the standard deviation.
A general trend is observed where $A_{\rm spot}$ appears to be
proportional to $\Delta F/F_{\rm av}$.
}\label{fig: A_dF_models_binned}\end{figure}

The mean follows a general trend such that
$A_{\rm spot} \propto \Delta F/F_{\rm av}$ (\Fig{fig: A_dF_models_binned}).
Of greater significance here is the standard deviation $\sigma$, which shows the
spread of the data for the various inclination angles and realizations.
We find that $\sigma$ is comparable to the slope of the general trend.
Together with the apparent trend,
we conclude that $\Delta F/F_{\rm av}$ is a useful proxy for
$A_{\rm spot}$.

\section{Flare Energy}

\subsection{Rotation Rate Affecting Flare Energy}

\begin{figure}[t!]\begin{center}
\includegraphics[width=\columnwidth]{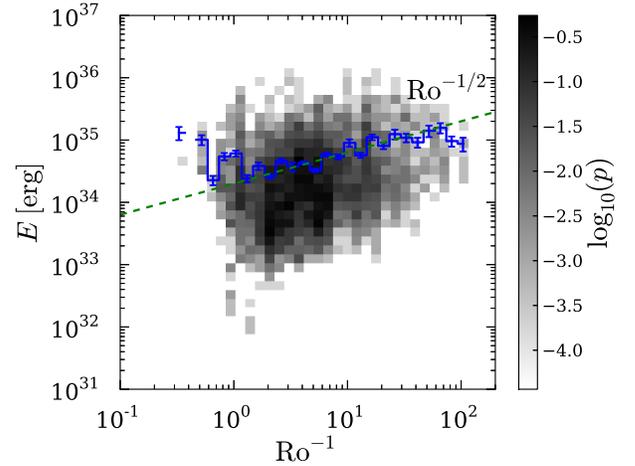}
\end{center}\caption[]{
Binned average for the flare energy in given intervals of $\Ro^{-1}$
(1D histogram) together with a fit.
The color mapping shows the relative population $p$ of flares within
intervals of $\Ro^{-1}$ and flare energy.
From the fit, it becomes clear that more energetic
flares can be expected from faster rotating stars.
}\label{fig: Ro-1_Ef_hist_mixed}\end{figure}

With increasing values of $\Ro^{-1}$, we first observe an increase and then a
decrease in $\nu_{\rm tot}\tau$ (\Fig{fig: nuTau_vs_Ro-1_mixed}).
We test whether or not the individual flare energy depends on $\Ro^{-1}$ using
a data set containing 6830 flares of 795 stars.
We know the Rossby number and brightness variation for 753 of those stars
for a total of 6568 flare events.
In \Fig{fig: Ro-1_Ef_hist_mixed}, we plot the relative flare population for
intervals of $\Ro^{-1}$ and flare energy $E$, and overplot a 1D
average.
Moving from slow to fast rotators, the average flare energy clearly systematically increases.
We can also identify a power law with an exponent of $1/2$.
However, the statistical significance is poor.

\subsection{Flare Energy Distribution}

\begin{figure}[t!]\begin{center}
\includegraphics[width=\columnwidth]{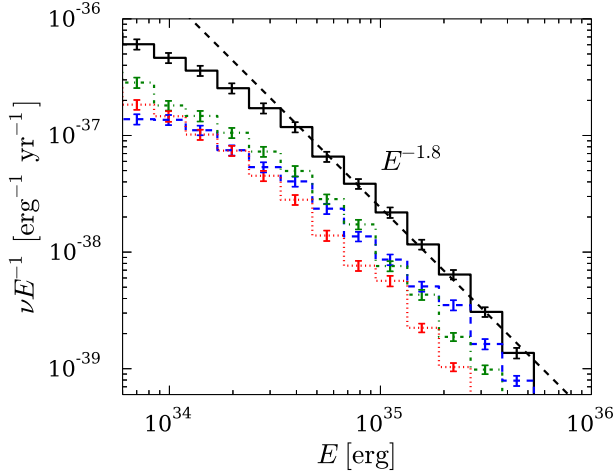}
\end{center}\caption[]{
Frequency distribution for the flare energy including all superflaring
stars (solid back line), stars with $\Ro^{-1} > 7$ (blue dashed line),
with $7 \ge \Ro^{-1} > 2.5$ (green dash-dotted line), and $\Ro^{-1} \le 2.5$
(red dotted line).
}\label{fig: nuE-1_vs_E}\end{figure}

\cite{Maehara2012Natur} and \cite{Shibayama-Maehara-2013-209-1-ApJS}
determined a power-law behavior for the dependence of flare frequency on flare energy.
The dependence was determined to be
proportional to $E^{-2.3}$ with an error of $\pm0.3$ for the slope.
By using a data set containing 6830 flares from 795 stars, we determine
the total number of flares within a given energy range and reproduce
their power-law behavior.
From our analysis, we find
$n_{\rm flares}(E_{\rm flare}) \times E_{\rm flare}^{-1} \propto
E_{\rm flare}^{-1.8}$, which is
comparable to previous findings (\Fig{fig: nuE-1_vs_E}).
We also check whether or not there is a different behavior for different ranges of
the Rossby number for $\Ro^{-1} > 7$, $7 \ge \Ro^{-1} > 2.5$,
and $\Ro^{-1} \le 2.5$, but we find no significant deviation.

\section{Interpretation and Modeling}

The above results show that while stellar activity is correlated
with $\Ro^{-1}$, flare energy is only poorly correlated
and shows significant scatter in this relation;
see \Fig{fig: Ro-1_Ef_hist_mixed}.
Whether or not this agrees quantitatively with the dynamo predictions
can be assessed through numerical simulations.

In turbulent dynamos, magnetic energy is distributed over a broad
range of scales.
The magnetic field seen in the solar cycle corresponds only to
the lowest wavenumbers of the magnetic energy spectrum.
However, the remaining part of the spectrum is quite independent
of the cycle and, presumably, also of the occurrence of grand minima.
Furthermore, we recall that evidence from $^{10}$Be isotope measurements
in the Greenland ice cores \citep{1998SoPh..181..237B} revealed cyclic
activity even during the Maunder minimum.
Therefore, even cycles themselves are independent of the overall activity state
of the system, be it in a grand minimum or grand maximum.
However, because the magnetic field remains highly turbulent,
involving all scales,
we must expect a certain level of fluctuations, which becomes
more intense toward smaller scales.

In hydromagnetic turbulence, magnetic dissipation is proportional to
the square of the current density, which is characteristic of the
smallest scales in the spectrum.
To further illustrate this, let us now consider a simple turbulent dynamo
exhibiting cyclic variability.
Crucial ingredients of such a dynamo are shear and helical turbulence.
This can easily be represented in a simulation of helically forced turbulence
with linear shear and shearing-periodic boundary conditions.
Such models have been studied extensively by \cite{KapylaBrandenb2009ApJ}
for different
values of the scale separation ratio $\kf/k_1$, where $\kf$ is the
forcing wavenumber and $k_1$ is the lowest wavenumber that fits into the
Cartesian domain of size $L^3$, so $k_1=2\pi/L$.
Applications to long-term variability have been studied by
\cite{BrandenburgGuerrero2012IAU286}
for small values of $\kf/k_1$ of 1.5 and 2.2, and different
shear parameters, $\Sh=S/\urms\kf$, where $S$ is the shear rate
characterizing the strength of the linear shear flow $\meanUU=(0,Sx,0)$
for a given rms velocity $\urms$ of the turbulence.
Applied to stellar differential rotation, the $y$ direction corresponds
to the toroidal direction and the $x$ direction to radius,
so that the $z$ direction corresponds to the latitude (i.e., negative colatitude).
We adopt negative values of $S$, which corresponds to the negative radial
shear in the near-surface shear layer.
In the solar dynamo, this layer may be important
in ``shaping'' the dynamo wave toward the equator
\citep{Brandenburg2005ApJ625,KosovichevPipinZhao2013}.

\begin{figure}[t!]\begin{center}
\includegraphics[width=\columnwidth]{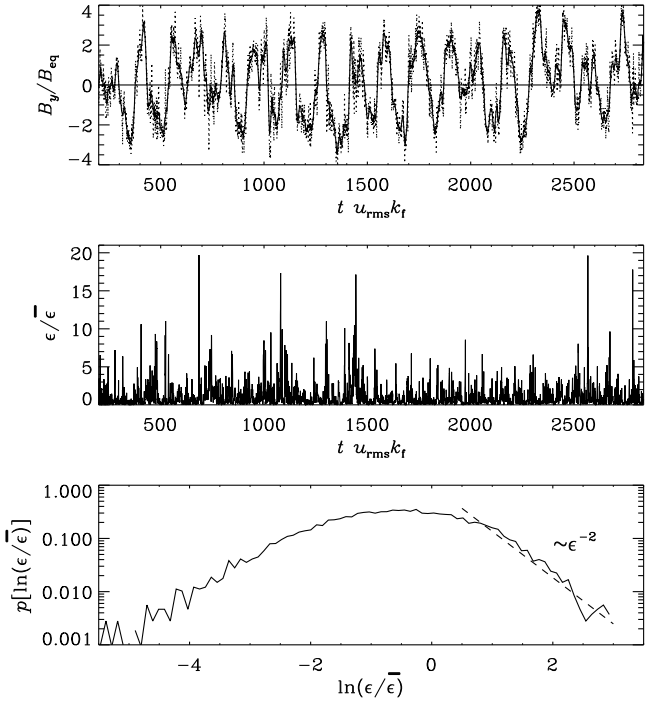}
\end{center}\caption[]{
Upper panel: time dependence of the toroidal magnetic field $B_y$,
normalized by the equipartition
field strength $\Beq$ (dotted line) compared with its temporally smoothed
evolution (solid line) for a model with $\kf/k_1=2.2$, $\Sh=-2.3$, and $D=27$.
Middle panel: local energy dissipation $\epsilon$ normalized by its averaged
value $\overline{\epsilon}$.
Lower panel: probability density function of $\epsilon/\overline{\epsilon}$.
The dashed line indicates a $p\propto\epsilon^{-2}$ scaling.
}\label{fig: pstat}\end{figure}

In the following, we consider the model of
\cite{BrandenburgGuerrero2012IAU286} for $\kf/k_1=2.2$
and five values of $\Sh$.
In particular, we study in detail the statistics of local magnetic
dissipation for such a model; see \Fig{fig: pstat} where we compare the local
evolution of the toroidal magnetic field $B_y$ (which shows cyclic behavior) with
that of the local magnetic dissipation rate, $\epsilon\propto\JJ^2$
(which shows no clear cycles).
Here, $\JJ=\nab\times\BB/\mu_0$ is the current density,
$\BB$ is the magnetic field, and $\mu_0$ is the vacuum permeability.
Note that in the present model where the magnetic Reynolds number is
only about 60, the maximum magnetic dissipation is more than 20 times
its average.
In turbulence theory, the local dissipation statistics is known to obey
a log-Poisson distribution for low local dissipation \citep{Dubrulle1994PRL},
but it is likely to show power-law behavior for high local dissipation
\citep{Gledzer1996PhFl}.
Our present results are roughly compatible with this; see \Fig{fig: pstat},
which shows $p(\ln\epsilon)\sim\epsilon^{-2}$.

To compare the resulting statistics for energy dissipation in the model with the
statistics of the flare energies of stars shown in \Fig{fig: Ro-1_Ef_hist_mixed},
we now consider models for different shear parameters $\Sh$.
The strength of the resulting large-scale dynamo is characterized by
the dynamo number $D$.
For $\alpha$--shear dynamos, $D$ is given by the product of two dynamo
numbers, $D=C_\alpha C_S$, where $C_\alpha=\alpha/\etat k_1$ measures
the relative strength of the kinetic helicity, and $C_S=S/\etat k_1^2$
measures the strength of the shear relative to turbulent diffusive
effects characterized by the turbulent magnetic diffusivity,
$\etat\approx\tau\urms^2/3$, where $\tau$ is the correlation time.
Estimating the $\alpha$ effect as $\alpha=-\tau\overline{\oo\cdot\uu}/3$
\citep{MoffattBook1978}, where $\overline{\oo\cdot\uu}$ is the kinetic
helicity with $\oo=\nab\times\uu$ being the vorticity of the flow $\uu$,
and estimating $\overline{\oo\cdot\uu}\approx\kf\urms^2$ for fully helical
turbulence \citep{alpha2_periodic13}, we find $C_\alpha\approx-\kf/k_1$.
Here, the minus sign in the expression for $\alpha$ is due to
the fact that $\alpha$ is a negative multiple of the kinetic helicity
and that the helicity of the turbulent forcing is positive.
For the second dynamo number, we similarly estimate $C_S=3\,\Sh\,(\kf/k_1)^2$ 
\citep{BrandenburgGuerrero2012IAU286}.
For the model presented in \Fig{fig: pstat}, we have $C_S\approx-12$.
Since $C_\alpha<0$, we have $D>0$, which yields dynamo waves traveling
in the positive $z$ direction, i.e., toward the equator.
Since $C_\alpha=\alpha/\etat k_1\approx-2.2$, we have $D\approx27$, which
is nearly 14 times larger than the critical value $D_{\rm crit}=2$ for the
onset of $\alpha$--shear dynamos \citep{BrandenbSubramanianReview2005}.

\begin{figure}[t!]\begin{center}
\includegraphics[width=\columnwidth]{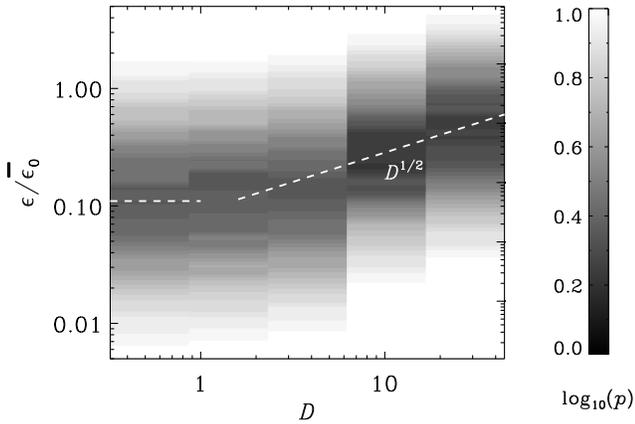}
\end{center}\caption[]{
Probability density function of $\ln(\epsilon/\overline{\epsilon}_0)$
for five bins of the dynamo number $D$.
}\label{fig: epsilon_dynamo_pdf_2d}\end{figure}

In \Fig{fig: epsilon_dynamo_pdf_2d}, we show the probability
density function of the normalized dissipation energy
$\ln(\epsilon/\overline{\epsilon}_0)$ and dynamo number, $D$,
where $\overline{\epsilon}_0$ is the average kinetic energy
input to the dynamo.
For $D<D_{\rm crit}$, the dynamo is just a small-scale dynamo,
where the median of the dissipation energy is independent of $D$.
For larger values of $D$, the median shows a mild increase
proportional to $D^{1/2}$, which is reminiscent of the increase
of the median of flare energies seen in \Fig{fig: Ro-1_Ef_hist_mixed}.
However, it is not clear how $D$ is related to $\Ro^{-1}$, but if
they were proportional to each other, then the two graphs would indeed
be in quantitative agreement with each other.
Furthermore, there is considerable scatter by about one dex,
and it might be even stronger for the flare energies seen
in \Fig{fig: Ro-1_Ef_hist_mixed}, where for a given value of
$\Ro^{-1}$ there can be significant variation.

\section{Conclusions}

In this paper, we have searched
for conditions under which flares with total energies above
$5\times10^{34}{\rm erg}$ occur.
We have used data from an extended superflare catalog
which had been derived from \textit{Kepler} data \citep{Maehara2012Natur}.
The stars are G-, K- and M-type stars.
Of those, only two are binary systems \citep{Matijevi2012AJ}.
There is no evidence for ``hot Jupiters'' orbiting the stars,
which makes our findings applicable to the Sun.
Given the similarity of the systems, we confirm the earlier findings of
\cite{Maehara2012Natur}, \cite{Shibata2012}, and
\cite{Shibayama-Maehara-2013-209-1-ApJS} that there is no need for any
external influence, which could affect the magnetic field in the corona,
as proposed by \cite{Rubenstein2000ApJ}.

The two important quantities we found were the effective temperature
and the inverse Rossby number, which is a
non-dimensional measure of the rotation rate.
Dynamo activity is known to decrease with the star's effective temperature
\citep{Kitchatinov-Olemskoy-2011-411-2-MNRAS} which then leads to less frequent
and less energetic flares.
We observe such a negative dependence for the monitored \textit{Kepler} stars
(\Fig{fig: nuTau_vs_Teff}).
From standard dynamo theory, we known that dynamo activity increases with
the rotation frequency \citep[e.g.,][]{Karak2014ApJ}.
In \Fig{fig: Ro-1_df_hist_mixed} (upper panel), we find this behavior for the observed stars
where we take the relative flux variation as a proxy for the starspot
coverage and magnetic activity.

Statistics from superflaring stars can be deceiving, as we observe
two very different results for the superflare occurrence rate
dependent on the Rossby number, based on whether or not
non-flaring stars are taken into account.
Using only superflare stars leads to no significant dependence of
the occurrence rate on the rotation rate (\Fig{fig: nuTau_vs_Ro-1_mixed}).
This is counterintuitive, since increased rotation should enhance the dynamo.
By including all of the observed stars, the average occurrence rate
changes due to the number of non-superflaring stars within that
bin (\Fig{fig: nuTau_vs_Ro-1_mixed}).
That way we obtain two power laws for $\nu_{\rm tot}\tau$ with the
powers $-2$ for $\Ro^{-1} \lesssim 10$ and $1$ for
$\Ro^{-1} \gtrsim 10$.
This finding is in agreement with \cite{Shibayama-Maehara-2013-209-1-ApJS},
who found higher superflare rates for fast rotating stars.

Observational bias arising from random angles between the observer--star
axis and its rotation axis is a considerable effect.
From the synthetic light curve measurements for our model
stars, we see a general trend as well as a significant spread
(\Fig{fig: A_dF_models_binned}).
We conclude that the inference of spot coverage from brightness variation
is valid, although it contains some uncertainties.

Flare energies are strongly connected with the rotation rate
(\Fig{fig: Ro-1_Ef_hist_mixed}).
This is expected from dynamo theory, as the increase in
magnetic energy is positively affected by the rotation rate.
The increased dynamo action leads to a higher coverage of spots
(\Fig{fig: Ro-1_df_hist_mixed}), and possibly to a higher number of
large spots.
Those large spots can store larger amounts of magnetic energy which
then leads to more energetic flares
(\Fig{fig: Ro-1_Ef_hist_mixed}, color mapping).

Our simulations of a standard dynamo with helically forced turbulence
clearly show a characteristic dependence for the Ohmic dissipation rate
(\Fig{fig: pstat}).
This $E^{-2}$ dependence should be compared to $E^{-2.3}$, found by
\cite{Maehara2012Natur} and \Fig{fig: nuE-1_vs_E}.
This shows that dissipation follows a power-law behavior.
Since the flares originated from such dissipations, this
explains the power-law behavior for the flare energy.
Exponential tails in the distribution of energy dissipation imply
that there is a considerable chance that an extreme dissipation event
or superflare could occur in a system whose average activity level
is comparatively low.

\acknowledgments

We thank the anonymous referee for useful and constructive comments.
We acknowledge the hospitality of Kyoto University
where large parts of this work were performed.
Computing resources
provided by the Swedish National Allocations
Committee at the Center for Parallel Computers at
the Royal Institute of Technology in Stockholm, and
the High Performance Computing Center North in Ume{\aa}.
This work was supported in part by
the European Research Council under the AstroDyn
Research Project No.\ 227952 and the Swedish Research
Council under grants 621-2011-5076 and 2012-5797,
as well as the Research Council of Norway
under the FRINATEK grant 231444.
A.H.\ is supported by KAKENHI Grant--in--Aid for Young Scientists
(B) 25800108.
H.M.\ is supported by a KAKENHI Grant--in--Aid for Young Scientists (B) 26800096
and K.S.\ is supported by the Grant--in--Aid from the Ministry of 
Education, Culture, Sports, Science, and Technology of Japan (B) 25287039.
\textit{Kepler} was selected as the 10th Discovery mission.
Funding for this  mission is provided by the NASA Science Mission Directorate.
The data presented in this paper were obtained from the Multimission Archive at 
STScI.

\bibliography{references}

\begin{thebibliography}{46}
\expandafter\ifx\csname natexlab\endcsname\relax\def\natexlab#1{#1}\fi

\bibitem[{{Barnes} {et~al.}(2005){Barnes}, {Collier Cameron}, {Donati},
  {James}, {Marsden}, \& {Petit}}]{Barnes-CollierCameron-2005-357-L1-MNRAS}
{Barnes}, J.~R., {Collier Cameron}, A., {Donati}, J.-F., {James}, D.~J.,
  {Marsden}, S.~C., \& {Petit}, P. 2005, \mnras, 357, L1

\bibitem[{{Beer} {et~al.}(1998){Beer}, {Tobias}, \&
  {Weiss}}]{1998SoPh..181..237B}
{Beer}, J., {Tobias}, S., \& {Weiss}, N. 1998, SoPh, 181, 237

\bibitem[{{Brandenburg}(2005)}]{Brandenburg2005ApJ625}
{Brandenburg}, A. 2005, \apj, 625, 539

\bibitem[{{Brandenburg} \& {Guerrero}(2012)}]{BrandenburgGuerrero2012IAU286}
{Brandenburg}, A., \& {Guerrero}, G. 2012, in IAU Symposium, Vol. 286,
  Comparative Magnetic Minima: Characterizing Quiet Times in the Sun and Stars,
  ed. C.~H. {Mandrini} \& D.~F. {Webb} (Cambridge University Press), 37

\bibitem[{{Brandenburg} {et~al.}(2012){Brandenburg}, {Sokoloff}, \&
  {Subramanian}}]{Brandenburg2012SSR}
{Brandenburg}, A., {Sokoloff}, D., \& {Subramanian}, K. 2012, SSRv, 169, 123

\bibitem[{{Brandenburg} \& {Subramanian}(2005)}]{BrandenbSubramanianReview2005}
{Brandenburg}, A., \& {Subramanian}, K. 2005, PhR, 417, 1

\bibitem[{Candelaresi \& Brandenburg(2013)}]{alpha2_periodic13}
Candelaresi, S., \& Brandenburg, A. 2013, PhRvE, 87, 043104

\bibitem[{{Carrington}(1859)}]{Carrington1859MNRAS}
{Carrington}, R.~C. 1859, \mnras, 20, 13

\bibitem[{{Choudhuri} {et~al.}(1995){Choudhuri}, {Sch\"ussler}, \&
  {Dikpati}}]{Choudhuri1995AA}
{Choudhuri}, A.~R., {Sch\"ussler}, M., \& {Dikpati}, M. 1995, \aap, 303, L29

\bibitem[{{Dikpati} \& {Charbonneau}(1999)}]{Dikpati1999ApJ}
{Dikpati}, M., \& {Charbonneau}, P. 1999, \apj, 518, 508

\bibitem[{{Dubrulle}(1994)}]{Dubrulle1994PRL}
{Dubrulle}, B. 1994, PhRvL, 73, 959

\bibitem[{Fr\"ohlich \& Lean(1998)}]{Frohlich-Lean-1998-25-23-GeoRL}
Fr\"ohlich, C., \& Lean, J. 1998, GeoRL, 25, 4377

\bibitem[{{Gledzer} {et~al.}(1996){Gledzer}, {Villermaux}, {Kahalerras}, \&
  {Gagne}}]{Gledzer1996PhFl}
{Gledzer}, E., {Villermaux}, E., {Kahalerras}, H., \& {Gagne}, Y. 1996, PhFl,
  8, 3367

\bibitem[{{Hodgson}(1859)}]{Hodgson1859MNRAS}
{Hodgson}, R. 1859, \mnras, 20, 15

\bibitem[{{K{\"a}pyl{\"a}} \& {Brandenburg}(2009)}]{KapylaBrandenb2009ApJ}
{K{\"a}pyl{\"a}}, P.~J., \& {Brandenburg}, A. 2009, \apj, 699, 1059

\bibitem[{{Karak} {et~al.}(2014){Karak}, {Kitchatinov}, \&
  {Choudhuri}}]{Karak2014ApJ}
{Karak}, B.~B., {Kitchatinov}, L.~L., \& {Choudhuri}, A.~R. 2014, \apj, 791, 59

\bibitem[{Kitchatinov \&
  Olemskoy(2011)}]{Kitchatinov-Olemskoy-2011-411-2-MNRAS}
Kitchatinov, L.~L., \& Olemskoy, S.~V. 2011, \mnras, 411, 1059

\bibitem[{{Koch} {et~al.}(2010){Koch}, {Borucki}, {Basri}, {Batalha}, {Brown},
  {Caldwell}, {Christensen-Dalsgaard}, {Cochran}, {DeVore}, {Dunham},
  {Gautier}, {Geary}, {Gilliland}, {Gould}, {Jenkins}, {Kondo}, {Latham},
  {Lissauer}, {Marcy}, {Monet}, {Sasselov}, {Boss}, {Brownlee}, {Caldwell},
  {Dupree}, {Howell}, {Kjeldsen}, {Meibom}, {Morrison}, {Owen}, {Reitsema},
  {Tarter}, {Bryson}, {Dotson}, {Gazis}, {Haas}, {Kolodziejczak}, {Rowe}, {Van
  Cleve}, {Allen}, {Chandrasekaran}, {Clarke}, {Li}, {Quintana}, {Tenenbaum},
  {Twicken}, \& {Wu}}]{KochKeplerMission2010}
{Koch}, D.~G., {et~al.} 2010, ApJL, 713, L79

\bibitem[{{Kosovichev} {et~al.}(2013){Kosovichev}, {Pipin}, \&
  {Zhao}}]{KosovichevPipinZhao2013}
{Kosovichev}, A.~G., {Pipin}, V.~V., \& {Zhao}, J. 2013, in ASP Conf.\ Ser.,
  Vol. 479, Progress in Physics of the Sun and Stars, ed. H.~{Shibahashi} \&
  A.~E. {Lynas-Gray} (San Francisco, CA: ASP), 395

\bibitem[{{Kovari} \& {Bartus}(1997)}]{Kovari-Bartus-1997-323-801-AA}
{Kovari}, Z., \& {Bartus}, J. 1997, \aap, 323, 801

\bibitem[{{Maehara} {et~al.}(2012){Maehara}, {Shibayama}, {Notsu}, {Notsu},
  {Nagao}, {Kusaba}, {Honda}, {Nogami}, \& {Shibata}}]{Maehara2012Natur}
{Maehara}, H., {et~al.} 2012, Natur, 485, 478

\bibitem[{Malanushenko {et~al.}(2014)Malanushenko, Schrijver, DeRosa, \&
  Wheatland}]{Malanushenko-Schrijver-2014-783-2-ApJ}
Malanushenko, A., Schrijver, C.~J., DeRosa, M.~L., \& Wheatland, M.~S. 2014,
  \apj, 783, 102

\bibitem[{{Matijevi{\v c}} {et~al.}(2012){Matijevi{\v c}}, {Pr{\v s}a},
  {Orosz}, {Welsh}, {Bloemen}, \& {Barclay}}]{Matijevi2012AJ}
{Matijevi{\v c}}, G., {Pr{\v s}a}, A., {Orosz}, J.~A., {Welsh}, W.~F.,
  {Bloemen}, S., \& {Barclay}, T. 2012, \aj, 143, 123

\bibitem[{Moffatt(1978)}]{MoffattBook1978}
Moffatt, H.~K. 1978, {Magnetic Field Generation in Electrically Conducting
  Fluids} (Cambridge: Cambridge Univ. Press)

\bibitem[{{Nandy} \& {Choudhuri}(2002)}]{Nandy2002Science}
{Nandy}, D., \& {Choudhuri}, A.~R. 2002, Sci, 296, 1671

\bibitem[{{Nogami} {et~al.}(2014){Nogami}, {Notsu}, {Honda}, {Maehara},
  {Notsu}, {Shibayama}, \& {Shibata}}]{Nogami-Notsu-2014-1402-arxiv}
{Nogami}, D., {Notsu}, Y., {Honda}, S., {Maehara}, H., {Notsu}, S.,
  {Shibayama}, T., \& {Shibata}, K. 2014, \pasj, 66, L4

\bibitem[{Notsu {et~al.}(2013)Notsu, Shibayama, Maehara, Notsu, Nagao, Honda,
  Ishii, Nogami, \& Shibata}]{Notsu2013ApJ}
Notsu, Y., {et~al.} 2013, \apj, 771, 127

\bibitem[{{Noyes} {et~al.}(1984){Noyes}, {Hartmann}, {Baliunas}, {Duncan}, \&
  {Vaughan}}]{Noyes1984ApJ}
{Noyes}, R.~W., {Hartmann}, L.~W., {Baliunas}, S.~L., {Duncan}, D.~K., \&
  {Vaughan}, A.~H. 1984, \apj, 279, 763

\bibitem[{{Pallavicini} {et~al.}(1981){Pallavicini}, {Golub}, {Rosner},
  {Vaiana}, {Ayres}, \& {Linsky}}]{Pallavicini1981}
{Pallavicini}, R., {Golub}, L., {Rosner}, R., {Vaiana}, G.~S., {Ayres}, T., \&
  {Linsky}, J.~L. 1981, \apj, 248, 279

\bibitem[{{Pizzolato} {et~al.}(2003){Pizzolato}, {Maggio}, {Micela},
  {Sciortino}, \& {Ventura}}]{Pizzolato2003}
{Pizzolato}, N., {Maggio}, A., {Micela}, G., {Sciortino}, S., \& {Ventura}, P.
  2003, \aap, 397, 147

\bibitem[{{Pouquet} {et~al.}(1976){Pouquet}, {Frisch}, \&
  {L\'eorat}}]{PouquetFrischLeorat1976JFM}
{Pouquet}, A., {Frisch}, U., \& {L\'eorat}, J. 1976, JFM, 77, 321

\bibitem[{{Reinhold} {et~al.}(2013){Reinhold}, {Reiners}, \&
  {Basri}}]{Reinhold-Reiners-2013-560-A4-AA}
{Reinhold}, T., {Reiners}, A., \& {Basri}, G. 2013, A\&A, 560, A4

\bibitem[{{Rubenstein} \& {Schaefer}(2000)}]{Rubenstein2000ApJ}
{Rubenstein}, E.~P., \& {Schaefer}, B.~E. 2000, \apj, 529, 1031

\bibitem[{{Sammis} {et~al.}(2000){Sammis}, {Tang}, \& {Zirin}}]{Sammis2000}
{Sammis}, I., {Tang}, F., \& {Zirin}, H. 2000, \apj, 540, 583

\bibitem[{{Schaefer} {et~al.}(2000){Schaefer}, {King}, \&
  {Deliyannis}}]{Schaefer2000ApJ}
{Schaefer}, B.~E., {King}, J.~R., \& {Deliyannis}, C.~P. 2000, \apj, 529, 1026

\bibitem[{{Schrijver}(2009)}]{Schrijver2009AdSpR43}
{Schrijver}, C.~J. 2009, AdSpR, 43, 739

\bibitem[{Shibata {et~al.}(2013)Shibata, Isobe, Hillier, Choudhuri, Maehara,
  Ishii, Shibayama, Notsu, Notsu, Nagao, Honda, \& Nogami}]{Shibata2012}
Shibata, K., {et~al.} 2013, \pasj, 65, 49

\bibitem[{Shibayama {et~al.}(2013)Shibayama, Maehara, Notsu, Notsu, Nagao,
  Honda, Ishii, Nogami, \& Shibata}]{Shibayama-Maehara-2013-209-1-ApJS}
Shibayama, T., {et~al.} 2013, \apjs, 209, 5

\bibitem[{Steenbeck {et~al.}(1966)Steenbeck, Krause, \&
  R{\"a}dler}]{steenbecketal66}
Steenbeck, M., Krause, F., \& R{\"a}dler, K.-H. 1966, ZNatA, 21, 369

\bibitem[{{Su} {et~al.}(2013){Su}, {Veronig}, {Holman}, {Dennis}, {Wang},
  {Temmer}, \& {Gan}}]{Su-Veronig-2013-9-489-NatPh}
{Su}, Y., {Veronig}, A.~M., {Holman}, G.~D., {Dennis}, B.~R., {Wang}, T.,
  {Temmer}, M., \& {Gan}, W. 2013, NatPh, 9, 489

\bibitem[{{Veltri} {et~al.}(2005){Veltri}, {Nigro}, {Malara}, {Carbone}, \&
  {Mangeney}}]{Veltri2005NPGeo}
{Veltri}, P., {Nigro}, G., {Malara}, F., {Carbone}, V., \& {Mangeney}, A. 2005,
  NPGeo, 12, 245

\bibitem[{{Vilhu}(1984)}]{Vilhu1984AA}
{Vilhu}, O. 1984, \aap, 133, 117

\bibitem[{{Walter}(1982)}]{Walter1981}
{Walter}, F.~M. 1982, \apj, 253, 745

\bibitem[{{Willson} {et~al.}(1981){Willson}, {Gulkis}, {Janssen}, {Hudson}, \&
  {Chapman}}]{Willson-Gulkis-1981-211-700-Sci}
{Willson}, R.~C., {Gulkis}, S., {Janssen}, M., {Hudson}, H.~S., \& {Chapman},
  G.~A. 1981, Sci, 211, 700

\bibitem[{{Willson} \& {Hudson}(1981)}]{Willson-Hudson-1981-244-L185-ApJL}
{Willson}, R.~C., \& {Hudson}, H.~S. 1981, ApJL, 244, L185

\bibitem[{{Wright} {et~al.}(2011){Wright}, {Drake}, {Mamajek}, \&
  {Henry}}]{Wright2011ApJ}
{Wright}, N.~J., {Drake}, J.~J., {Mamajek}, E.~E., \& {Henry}, G.~W. 2011,
  \apj, 743, 48

\end{thebibliography}
\end{document}